\documentclass[prl,superscriptaddress,twocolumn,showpacs,amsmath,amssymb]{revtex4}

\usepackage{epsfig}
\usepackage{dcolumn}
\usepackage{bm}
\usepackage{ulem}

\usepackage{graphicx}

\include{include}

\newcommand{\MetRaw}{\mbox{$E\kern-0.60em\raise0.10ex\hbox{/}_{T}^{\rm raw}$}}
\newcommand{\MetVec}{\mbox{$\vec E\kern-0.60em\raise0.10ex\hbox{/}_{T}$}}
\newcommand{\MetVecRaw}{\mbox{$\vec E\kern-0.60em\raise0.10ex\hbox{/}^{\rm raw}_{T}$}}
\newcommand{\Met}{\mbox{$E\kern-0.60em\raise0.10ex\hbox{/}_{T}$}}
\newcommand{\MetSpec}{\mbox{$E\kern-0.60em\raise0.10ex\hbox{/}_{T}^{spec}$}}
\newcommand{\MetRel}{\mbox{$E\kern-0.60em\raise0.10ex\hbox{/}_{T, rel}$}}

\begin{document}

\title{Measurement of the $WZ$ Cross Section and Triple Gauge Couplings 
in $p \bar p$ Collisions at $\sqrt{s} = 1.96$ TeV}
\date{\today}

\affiliation{Institute of Physics, Academia Sinica, Taipei, Taiwan 11529, Republic of China}
\affiliation{Argonne National Laboratory, Argonne, Illinois 60439, USA}
\affiliation{University of Athens, 157 71 Athens, Greece}
\affiliation{Institut de Fisica d'Altes Energies, ICREA, Universitat Autonoma de Barcelona, E-08193, Bellaterra (Barcelona), Spain}
\affiliation{Baylor University, Waco, Texas 76798, USA}
\affiliation{Istituto Nazionale di Fisica Nucleare Bologna, $^{ee}$University of Bologna, I-40127 Bologna, Italy}
\affiliation{University of California, Davis, Davis, California 95616, USA}
\affiliation{University of California, Los Angeles, Los Angeles, California 90024, USA}
\affiliation{Instituto de Fisica de Cantabria, CSIC-University of Cantabria, 39005 Santander, Spain}
\affiliation{Carnegie Mellon University, Pittsburgh, Pennsylvania 15213, USA}
\affiliation{Enrico Fermi Institute, University of Chicago, Chicago, Illinois 60637, USA}
\affiliation{Comenius University, 842 48 Bratislava, Slovakia; Institute of Experimental Physics, 040 01 Kosice, Slovakia}
\affiliation{Joint Institute for Nuclear Research, RU-141980 Dubna, Russia}
\affiliation{Duke University, Durham, North Carolina 27708, USA}
\affiliation{Fermi National Accelerator Laboratory, Batavia, Illinois 60510, USA}
\affiliation{University of Florida, Gainesville, Florida 32611, USA}
\affiliation{Laboratori Nazionali di Frascati, Istituto Nazionale di Fisica Nucleare, I-00044 Frascati, Italy}
\affiliation{University of Geneva, CH-1211 Geneva 4, Switzerland}
\affiliation{Glasgow University, Glasgow G12 8QQ, United Kingdom}
\affiliation{Harvard University, Cambridge, Massachusetts 02138, USA}
\affiliation{Division of High Energy Physics, Department of Physics, University of Helsinki and Helsinki Institute of Physics, FIN-00014, Helsinki, Finland}
\affiliation{University of Illinois, Urbana, Illinois 61801, USA}
\affiliation{The Johns Hopkins University, Baltimore, Maryland 21218, USA}
\affiliation{Institut f\"{u}r Experimentelle Kernphysik, Karlsruhe Institute of Technology, D-76131 Karlsruhe, Germany}
\affiliation{Center for High Energy Physics: Kyungpook National University, Daegu 702-701, Korea; Seoul National University, Seoul 151-742, Korea; Sungkyunkwan University, Suwon 440-746, Korea; Korea Institute of Science and Technology Information, Daejeon 305-806, Korea; Chonnam National University, Gwangju 500-757, Korea; Chonbuk National University, Jeonju 561-756, Korea}
\affiliation{Ernest Orlando Lawrence Berkeley National Laboratory, Berkeley, California 94720, USA}
\affiliation{University of Liverpool, Liverpool L69 7ZE, United Kingdom}
\affiliation{University College London, London WC1E 6BT, United Kingdom}
\affiliation{Centro de Investigaciones Energeticas Medioambientales y Tecnologicas, E-28040 Madrid, Spain}
\affiliation{Massachusetts Institute of Technology, Cambridge, Massachusetts 02139, USA}
\affiliation{Institute of Particle Physics: McGill University, Montr\'{e}al, Qu\'{e}bec, Canada H3A~2T8; Simon Fraser University, Burnaby, British Columbia, Canada V5A~1S6; University of Toronto, Toronto, Ontario, Canada M5S~1A7; and TRIUMF, Vancouver, British Columbia, Canada V6T~2A3}
\affiliation{University of Michigan, Ann Arbor, Michigan 48109, USA}
\affiliation{Michigan State University, East Lansing, Michigan 48824, USA}
\affiliation{Institution for Theoretical and Experimental Physics, ITEP, Moscow 117259, Russia}
\affiliation{University of New Mexico, Albuquerque, New Mexico 87131, USA}
\affiliation{The Ohio State University, Columbus, Ohio 43210, USA}
\affiliation{Okayama University, Okayama 700-8530, Japan}
\affiliation{Osaka City University, Osaka 588, Japan}
\affiliation{University of Oxford, Oxford OX1 3RH, United Kingdom}
\affiliation{Istituto Nazionale di Fisica Nucleare, Sezione di Padova-Trento, $^{ff}$University of Padova, I-35131 Padova, Italy}
\affiliation{University of Pennsylvania, Philadelphia, Pennsylvania 19104, USA}
\affiliation{Istituto Nazionale di Fisica Nucleare Pisa, $^{gg}$University of Pisa, $^{hh}$University of Siena and $^{ii}$Scuola Normale Superiore, I-56127 Pisa, Italy}
\affiliation{University of Pittsburgh, Pittsburgh, Pennsylvania 15260, USA}
\affiliation{Purdue University, West Lafayette, Indiana 47907, USA}
\affiliation{University of Rochester, Rochester, New York 14627, USA}
\affiliation{The Rockefeller University, New York, New York 10065, USA}
\affiliation{Istituto Nazionale di Fisica Nucleare, Sezione di Roma 1, $^{jj}$Sapienza Universit\`{a} di Roma, I-00185 Roma, Italy}
\affiliation{Rutgers University, Piscataway, New Jersey 08855, USA}
\affiliation{Texas A\&M University, College Station, Texas 77843, USA}
\affiliation{Istituto Nazionale di Fisica Nucleare Trieste/Udine, I-34100 Trieste, $^{kk}$University of Udine, I-33100 Udine, Italy}
\affiliation{University of Tsukuba, Tsukuba, Ibaraki 305, Japan}
\affiliation{Tufts University, Medford, Massachusetts 02155, USA}
\affiliation{University of Virginia, Charlottesville, Virginia 22906, USA}
\affiliation{Waseda University, Tokyo 169, Japan}
\affiliation{Wayne State University, Detroit, Michigan 48201, USA}
\affiliation{University of Wisconsin, Madison, Wisconsin 53706, USA}
\affiliation{Yale University, New Haven, Connecticut 06520, USA}

\author{T.~Aaltonen}
\affiliation{Division of High Energy Physics, Department of Physics, University of Helsinki and Helsinki Institute of Physics, FIN-00014, Helsinki, Finland}
\author{B.~\'{A}lvarez~Gonz\'{a}lez$^z$}
\affiliation{Instituto de Fisica de Cantabria, CSIC-University of Cantabria, 39005 Santander, Spain}
\author{S.~Amerio}
\affiliation{Istituto Nazionale di Fisica Nucleare, Sezione di Padova-Trento, $^{ff}$University of Padova, I-35131 Padova, Italy}
\author{D.~Amidei}
\affiliation{University of Michigan, Ann Arbor, Michigan 48109, USA}
\author{A.~Anastassov$^x$}
\affiliation{Fermi National Accelerator Laboratory, Batavia, Illinois 60510, USA}
\author{A.~Annovi}
\affiliation{Laboratori Nazionali di Frascati, Istituto Nazionale di Fisica Nucleare, I-00044 Frascati, Italy}
\author{J.~Antos}
\affiliation{Comenius University, 842 48 Bratislava, Slovakia; Institute of Experimental Physics, 040 01 Kosice, Slovakia}
\author{G.~Apollinari}
\affiliation{Fermi National Accelerator Laboratory, Batavia, Illinois 60510, USA}
\author{J.A.~Appel}
\affiliation{Fermi National Accelerator Laboratory, Batavia, Illinois 60510, USA}
\author{T.~Arisawa}
\affiliation{Waseda University, Tokyo 169, Japan}
% ...

\collaboration{CDF Collaboration\footnote{With visitors from
$^a$Istituto Nazionale di Fisica Nucleare, Sezione di Cagliari, 09042 Monserrato (Cagliari), Italy,
$^b$University of CA Irvine, Irvine, CA 92697, USA,
$^c$University of CA Santa Barbara, Santa Barbara, CA 93106, USA,
$^d$University of CA Santa Cruz, Santa Cruz, CA 95064, USA,
$^e$Institute of Physics, Academy of Sciences of the Czech Republic, Czech Republic,
$^f$CERN, CH-1211 Geneva, Switzerland,
$^g$Cornell University, Ithaca, NY 14853, USA,
$^h$University of Cyprus, Nicosia CY-1678, Cyprus,
$^i$Office of Science, U.S. Department of Energy, Washington, DC 20585, USA,
$^j$University College Dublin, Dublin 4, Ireland,
$^k$ETH, 8092 Zurich, Switzerland,
$^l$University of Fukui, Fukui City, Fukui Prefecture, Japan 910-0017,
$^m$Universidad Iberoamericana, Mexico D.F., Mexico,
$^n$University of Iowa, Iowa City, IA 52242, USA,
$^o$Kinki University, Higashi-Osaka City, Japan 577-8502,
$^p$Kansas State University, Manhattan, KS 66506, USA,
$^q$Ewha Womans University, Seoul, 120-750, Korea
$^r$University of Manchester, Manchester M13 9PL, United Kingdom,
$^s$Queen Mary, University of London, London, E1 4NS, United Kingdom,
$^t$University of Melbourne, Victoria 3010, Australia,
$^u$Muons, Inc., Batavia, IL 60510, USA,
$^v$Nagasaki Institute of Applied Science, Nagasaki, Japan,
$^w$National Research Nuclear University, Moscow, Russia,
$^x$Northwestern University, Evanston, IL 60208, USA,
$^y$University of Notre Dame, Notre Dame, IN 46556, USA,
$^z$Universidad de Oviedo, E-33007 Oviedo, Spain,
$^{aa}$CNRS-IN2P3, Paris, F-75205 France,
$^{bb}$Texas Tech University, Lubbock, TX 79609, USA,
$^{cc}$Universidad Tecnica Federico Santa Maria, 110v Valparaiso, Chile,
$^{dd}$Yarmouk University, Irbid 211-63, Jordan,
}}
\noaffiliation


\begin{abstract} 
This Letter describes the current most precise measurement of the $WZ$
production cross section as well as limits on
anomalous $WWZ$ couplings at a center-of-mass energy of 1.96 TeV in
proton-antiproton collisions for the Collider Detector at Fermilab (CDF).  
$WZ$ candidates are reconstructed
from decays containing three charged leptons and missing energy from a neutrino, where
the charged leptons are either electrons or muons.  Using data
collected by the CDF II detector (7.1 fb$^{-1}$ of integrated
luminosity), 63 candidate events are observed with the
expected background contributing $8 \pm 1$ events.  The measured
total cross section $\sigma (p \bar p \to WZ) =
3.93_{-0.53}^{+0.60}(\text{stat})_{-0.46}^{+0.59}(\text{syst}) $ pb
is in good agreement with the standard model prediction of $3.50\pm
0.21$.  The same sample is used to set limits on anomalous $WWZ$
couplings.

\end{abstract}


\maketitle

\normalsize


The measurement of $WZ$ production is an important test of the
standard model (SM) of particle physics.  $WZ$ pairs are produced in both
$s$-channel ($q\bar{q}\to W^{*}\to WZ$) and in $t$-channel
($q\bar{q}\to WZ$) interactions. 
The $WZ$ production is unique in that
the $s$-channel mode of production provides sensitivity to the $WWZ$
vertex which is governed by triple-gauge boson couplings (TGCs); the
presence of anomalous couplings~\cite{ATGC} could be an indication of
new physics at a higher mass scale leading to different rates and
kinematic distributions than predicted by the SM.  Furthermore, this
process is an essential background for Higgs boson
searches at particle colliders because the $WZ$ decay into leptons is the primary background
to high mass Higgs boson searches in the three-lepton signature, as
well as an important background process in two-lepton Higgs boson
analyses \cite{CDFHWW}.


This Letter reports a measurement of the $WZ$ production cross section
and limits on anomalous TGCs using a final state consisting of three
charged leptons and one neutrino in $p \bar p$ collision data
collected by the CDF II detector from 7.1 fb$^{-1}$ of integrated
luminosity.  The $WZ\to l\nu ll$ decay, where $l$ is an electron or muon 
and $\nu$ is a neutrino contributing missing energy, 
allows for the reconstruction of a variety of kinematic
quantities that are utilized to distinguish signal from background by
training a neural network \cite{NeuroBayes}.  The cross section is
then extracted from the output of the neural network by fitting the
shape with a maximum likelihood method.  Limits on anomalous TGCs are
set by analyzing the shape of the component of the momentum of the $Z$
boson that is transverse to the beamline as a discriminant.

The $SU(2)_L\times U(1)_Y$ part of the SM implies that $W$ and $Z$ weak
vector bosons may interact via trilinear or quartic vertices.
The $WZ$ cross section is proportional to the interaction coupling
strength predicted by the SM as $e\cot\theta_W$, where $e$ is the
electric charge and $\theta_W$ is the weak mixing angle.
We want to measure the cross section of the process $p \bar p\rightarrow WZ$, 
whose expected value in the limit of zero $W$ and $Z$ boson decay widths is 3.50 pb \cite{MCFM,MSTW2008}.
We therefore
measure the cross section by performing a likelihood fit to
a parameter representing the ratio of the measured to expected
$WZ$ cross section, to be discussed below.


The cross section of this process was first reported by CDF
with 1.1 fb$^{-1}$~\cite{Neubauer} in 2007. 
Subsequently, D\O\ reported with 1.0 fb$^{-1}$~\cite{D0:1000} in 2005, updated with 4.1
fb$^{-1}$~\cite{D0:4100} in 2007, and again with 8.6 fb$^{-1}$~\cite{D0:8600} in 2012.
Limits on anomalous TGCs were reported previously by 
LEP2~\cite{lep_tgc}, D\O\ ~\cite{D0atgc}, and recently by 
CMS~\cite{cms_tgc} and ATLAS~\cite{atlas_tgc}.
All results reported measurements consistent with the standard model.
This analysis describes CDF's most precise measurement of the $WZ$ cross section
and TGCs.

In the CDF II detector~\cite{CDFII}, a particle's direction is
characterized by the azimuthal angle $\phi$ and the pseudorapidity
$\eta = -\ln [ \tan (\theta/2) ]$, where $\theta$ is the polar angle
measured with respect to the proton beam direction.  The transverse energy $E_T$
is defined as $E \sin \theta$, where $E$ is the energy in the
electromagnetic (EM) and hadronic calorimeter towers associated with
a cluster of energy deposition.
The transverse momentum $p_T$ is the
particle's momentum component transverse to the beam line.  
The magnitude
of the $p_T$ for an electron is scaled according to the energy
measured in the calorimeter in order to account for momentum loss from final state radiation and
bremsstrahlung.  The missing transverse energy vector  $\MetVec$  is
defined as $-\sum_{i} E_T^i ~ \hat{n}_T^i$, where the index $i$ loops
over all towers of the calorimeter and $\hat{n}_T^i$ is the unit vector in
the transverse plane pointing from the interaction point to the energy
deposition in calorimeter tower $i$.  The $\MetVec$ is corrected for the $p_T$
of muons, which do not deposit all of their energy in the calorimeter,
and tracks that point to uninstrumented regions in the calorimeter.
The scalar missing transverse energy is defined as $|\MetVec|$ and denoted as $\Met$.
Strongly interacting partons produced in the $p \bar p$ collision
undergo fragmentation that results in highly collimated jets of
hadronic particles.  Jet candidates are reconstructed using the
calorimeter signals and are required to have 
$E_T > 15$ GeV and $|\eta| < 2.5$.
Isolated lepton candidates are accepted out to an $|\eta|$ of 2.0 for
electron candidates and $|\eta|$ of 1.0 for muon candidates.

The experimental signature for the decay $WZ\to l\nu ll$ is
reconstructed as three charged leptons (electrons or muons) and $\Met$
from the neutrino(s) that escaped undetected.  Events are also
detected if the $W$ or $Z$ decays to tau lepton(s) and those tau(s)
subsequently decay to detectable electrons or muons--these events are
considered part of the signal.
Consequently, events containing three charged leptons, not all with the same charge, are
selected from the data sample.  The online event triggering and
selection of lepton candidates are identical to those used in the
search for SM Higgs bosons decaying to two $W$ bosons at
CDF~\cite{CDFHWW}.  Our baseline event selection is to require the leading-lepton's 
$E_T$ (or $p_T$ for muons) 
to be above 20 GeV (GeV/$c$) to satisfy the trigger
requirements, while the second and third leptons are allowed to have
an $E_T$ ($p_T$) as low as $10$ GeV (GeV/$c$).  Additionally, because
the neutrino in the $W\rightarrow l\nu$ decay carries undetected
energy, the $WZ\rightarrow l\nu ll$ process tends to produce events
with higher missing energy than the background processes--aside
from $t\bar{t}$ whose $\Met$ distribution is also similarly high-valued, but
is a nearly negligible background.  We
therefore require $\Met > 25$ GeV.

Lastly, the dominant background remaining after these cuts is SM
$ZZ$ production.  This motivates two more cuts to require one and
only one $Z\rightarrow ll$ candidate in the event.  
We make a standard $Z$
boson identification cut by requiring events to have a pair of
same-flavor, opposite-signed leptons whose two-lepton mass falls
within a window of $\pm15$ GeV$/c^2$ around the $Z$ mass. This
removes most of the SM backgrounds with no $Z$ in the final state.
We note that this cut reduces $Z\gamma\rightarrow ll\gamma$ events because
the dilepton mass would not reconstruct back to the $Z$ when the $\gamma$ is
emitted by one of the two leptons.  In that case, the two-lepton mass
underestimates the $Z$ mass because the three-body ($ll\gamma$) mass
is what reconstructs the $Z$.

To further reduce the $ZZ\rightarrow llll$
background, we reject any event with an extra track with $p_T>8$ GeV$/c$,
thereby rejecting events that may have a fourth lepton that failed to
be identified.  This cut reduces the remaining $ZZ$ background by
$\sim 36\%$ while leaving the $WZ$ signal contribution essentially
unchanged.  Even so, $ZZ$ remains the primary background in this measurement.

There are several SM processes that
result in a similar final state to $WZ$ and are backgrounds in this
measurement.  The aforementioned $ZZ\to llll$ process appears as a background when
one of the four
leptons fails to be reconstructed by the detector.  This leaves three
reconstructed leptons with the one lepton failing reconstruction
providing the missing energy signature.  Drell-Yan events produced in
association with hadronic jets that mimic the signature of a third
lepton as well as Drell-Yan pairs produced with an associated photon that
converts to an electron-positron pair via interaction with the detector are also
significant backgrounds.  Lastly, top quark pair production
($t\bar{t}\rightarrow W^+ b W^- \bar{b}$) provides a minor
contribution to the background when one of the subsequent $b$-quark
jets mimics a lepton signature.  The sum of these four backgrounds is
quite small compared to the expected signal in the signal kinematics
region.

The background modeling--with the exception of the
$Z+$jets background--is MC simulated.  Events
from $WZ$, $ZZ$, and $t\bar{t}$ are simulated using the {\sc
pythia}~\cite{PYTHIA} generator.  The $Z\gamma$ background
is determined using the generator described in
Ref.~\cite{BAUR}.  The response of the CDF II detector is modeled
with a {\sc geant3}-based simulation~\cite{GEANT} program.  The expected
yields for each process are normalized to the cross sections
calculated at partial next-to next-to-leading order ($t \bar
t$~\cite{ttbarXsec}), next-to-leading order ($WZ$ and $ZZ$~\cite{MCFM}), or
leading-order with an estimated normalization correction to account for higher-orders
($Z\gamma$~\cite{BAUR}).  Efficiency corrections for the simulated
detector response to lepton candidates are determined using samples of
observed $Z \to l^+ l^-$ events.  
The $Z+$jets background
normalization is calculated using the probability that a hadronic jet will be
reconstructed as a lepton candidate (the same as is done in CDF's
$H\rightarrow WW$ search \cite{CDFHWW}), which is measured
in independent jet-triggered data samples.  
These probabilities are
applied to the jets in the $Z+$jets data sample to estimate the number
of such events that will pass the lepton identification and
signal selection criteria.  The expected signal and background
contributions are given in Table~\ref{tbl:WZExpected} along with the
observed number of events.

\begin{table}
\centering
\caption{Expected number of signal ($WZ$) and background events along with the
	total number of expected and observed events in the data.  Uncertainties 
	include all systematic uncertainties described in the text.}
\label{tbl:WZExpected}
\begin{tabular}{lrcr}\hline\hline
Process           &   \multicolumn{3}{c}{Events}   \\\hline
$ZZ$              &          3.6 &$\pm$&    0.5  \\
$Z+$jets          &          3.4 &$\pm$&    0.8  \\
$Z\gamma$         &          0.8 &$\pm$&    0.3  \\
$t \bar t$        &          0.1 &$\pm$&    0.04  \\\hline
Total background  &          7.9 &$\pm$&    1.0  \\\hline
$WZ$              &          47.4 &$\pm$&    4.8  \\\hline
Total expected    &          55.3 &$\pm$&    4.9  \\\hline
Data              &  \multicolumn{3}{c}{63}       \\\hline\hline
\end{tabular}
\end{table}

The dominant systematic uncertainties on the estimated contributions come from the luminosity 
measurement (6\%)~\cite{LUMI} and the simulated acceptances of the signal and background processes.
The acceptance uncertainty due to the parton distribution function modeling
ranges from 2.1\% to 2.7\% for the various processes.
A 10\% uncertainty is assigned to $WZ$ and $ZZ$ processes
for the kinematic differences between leading-order and higher-order calculations.
The cross section uncertainty is $6\%$ on the $ZZ$ process, $7\%$ on $t \bar t$, and
$5\%$ on $Z\gamma$.  The $Z\gamma$ process has another $20\%$ uncertainty that
accounts for possible mismodeling of the rate at which the $\gamma$ is
misidentified as a lepton.  
Similarly, there is a $25\%$ ($23\%$) uncertainty for $Z+$jets ($t \bar t$) for mismodeling
the rate at which light jets ($b$-jets) are misidentified as a lepton.
The uncertainty for the modeling of lepton identification is $2\%$ and of trigger efficiencies is $5.4\%$.
Lastly, uncertainties for overall rates for the modeling of jets accounts for $1.2\%$.

\begin{figure}
\centering
\includegraphics[width=\linewidth]{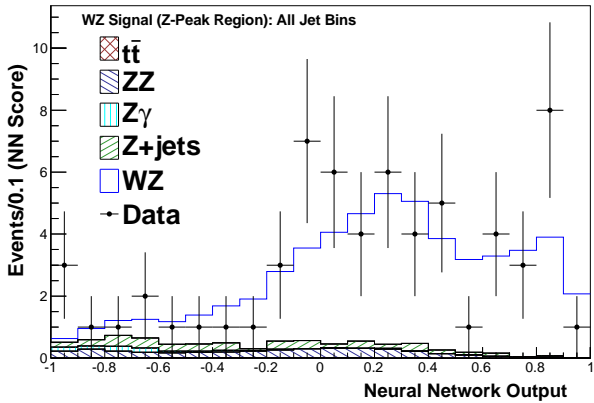}
\caption{The NN output for discriminating the $WZ$ signal events from
background processes within the selected signal sample.  Note that the
$t\bar{t}$ contribution is small enough to not be visible.  The processes are stacked.}
\label{fig:WZNN}
\end{figure}

Within the signal kinematic region, we seek to further isolate the signal from background by
utilizing a NeuroBayes neural network treatment \cite{NeuroBayes}.  
In general, the benefits of using a neural network (NN) over a simple counting experiment
are two-fold: it can better isolate the signal from the background and provide
a single distribution from which the cross section value can be extracted by fitting
the data to the shape of the expected physical processes.
We train a 
neural network with a combination of background events and simulated signal
events.  The input variables for the NN are
kinematic quantities selected to exploit differences between
signal and background distributions.  Starting with many quantities that show
relatively small differences in the distributions of backgrounds and signal, a
neural network will assign a numerical score whose distribution for
backgrounds and signal will be better separated than in any single input quantity alone.
The $\Met$ is a very useful input quantity for the NN because the $W\to l\nu$ 
decay in the signal yields a $\Met$ distribution with higher values than
the backgrounds.  Similarly, the azimuthal angle distribution between the $W$ lepton
and the $\Met$ is useful for distinguishing $WZ$ from the backgrounds
because they do not contain $W$ decays.
The total energy transverse to the beamline 
deposited by the $WZ$ decay compared to that of background
processes and lepton flavor combinations 
($eee$, $ee\mu$, $e\mu \text{ track}$, etc.) are also examples of NN input variables used.  Figure \ref{fig:WZNN}
shows the output of the NN treatment, with background-like events in simulation and data
trending toward a value of $-1$ while signal-like events trend towards $+1$.
Note that $t\bar{t}$ is represented in Fig. \ref{fig:WZNN}, but
has too small of a contribution to be visible.

The measured cross section for $WZ$ is extracted from the NN output in Fig.
\ref{fig:WZNN} with a binned maximum likelihood fit method.  The likelihood function is
formed from a product of Poisson probabilities for each bin in the NN output
and Gaussian constraints are applied 
corresponding to each systematic uncertainty:
\begin{equation}
\label{eqn:Likelihood}
\mathcal{L} = \left(\prod_i \frac{\mu_i^{n_i} e^{-\mu_i}}{n_i!} \right) \cdot \prod_c e^\frac{-S_c^2}{2}
\end{equation}
where $\mu_i$ is the total expectation in the $i$-th bin, $n_i$ is the 
number of data events in the $i$-th bin, and $S_c$ is a floating parameter associated
with the systematic uncertainty $c$.  The $\mu_i$ parameter is given by
\begin{equation}
\label{eqn:Mu}
\mu_i = \sum_k \left(\frac{\sigma^{\text{measured}}}{\sigma^{\text{expected}}}\right)_{k} \left[ \prod_c (1 + f^c_k S_c ) \right]  (N^{\text{expected}}_k)_i
\end{equation}
The sum $k$ is over the five processes (one signal and four background) that can contribute to events in bin $i$ and
$f^c_k$ is the fractional uncertainty for the process $k$ due to
the systematic uncertainty $c$.  
Some systematic uncertainties are common to more than
one process and so are correlated.  These correlations are accounted for in the
definition of $\mu_i$ through the $f^c_k$ parameters.  
The $(N^{\text{expected}}_k)_i$ is
the expected number of events from process $k$ in the $i$-th bin.
All the background processes are constrained to their SM expectations
by setting the proportion of measured to expected cross section to unity.  The likelihood is
then maximized with respect to the floating systematic
($S_c$) and cross section proportion 
$\left(\frac{\sigma^{\text{measured}}}{\sigma^{\text{expected}}}\right)_{WZ}$ parameters,
where $\sigma^{\text{expected}}$ is the expected signal cross section and
$\sigma^{\text{measured}}$  is the $WZ$ cross section ultimately measured from the data.
This method gives a measured value for the $WZ$ cross section of
$\sigma (p \bar p \to WZ) =
3.93_{-0.53}^{+0.60}(\text{stat})_{-0.46}^{+0.59}(\text{syst}) $ pb, which
is in good agreement with the aforementioned standard model prediction of $3.50\pm
0.21$\cite{MCFM,MSTW2008}.

\begin{figure}
\centering
\includegraphics[width=\linewidth]{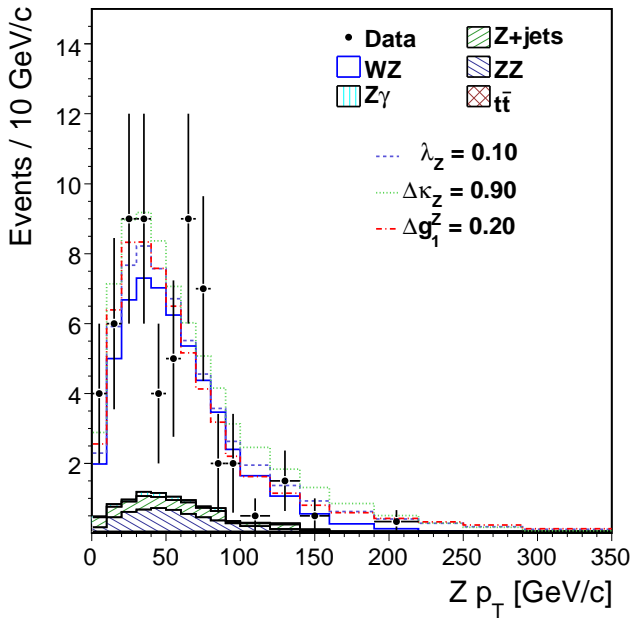}
\caption{The $Z$ $p_T$ distributions for data compared to the SM
  expectation for signal ($WZ$) and background. Also presented is how
  the signal expectation would change with the introduction of
  anomalous couplings near the observed limits.  The processes are stacked.}
\label{fig:Zpt}
\end{figure}

The shape and normalization of the $p_T$ spectrum of the $Z$ boson
(Fig. \ref{fig:Zpt}) are used to place limits on anomalous TGCs. The
most general modification of the $WWZ$ vertex preserving $C$ and
$P$ separately is parametrized by $\lambda_Z$, $g_1^Z$
and $\kappa_Z$~\cite{tgccutoff}. In the SM, $\lambda_Z = \Delta g_1^Z
= \Delta \kappa_Z = 0$ where $\Delta g_1^Z$ and $\Delta \kappa_Z$ are
used to denote the deviations of $g_1^Z$ and $\kappa_Z$ from their SM
values.
In general, the parameters $\lambda_z$, $\Delta g_1^Z$ and $\Delta
\kappa_Z$ can be functions of the invariant mass $\sqrt{\hat{s}}$ of the $WZ$ system. 
Non-zero values of $\lambda_Z$, $g_1^Z$ and
$\kappa_Z$ at large $\hat{s}$ violate unitarity. To avoid this, each coupling is modified by a form
factor $\alpha(\hat{s}) =
\frac{\alpha_0}{(1+\hat{s}/\Lambda^2)^2}$, where $\alpha_0$ 
is the unmodified coupling $\lambda_z$, $g_1^Z$ or $\kappa_Z$.


The likelihood of the $Z$ $p_T$ distribution for various anomalous TGC
models is used to set limits. The expected $Z$ $p_T$ distribution for
a given TGC before the effect of the detector response is obtained
using {\sc MCFM}~\cite{MCFM}. The detector acceptance and efficiency
are modeled by multiplying the {\sc MCFM} distribution by a $Z$ $p_T$-dependent factor.  
This factor is calculated using six different
simulated event samples generated at different TGC values with the
full detector response simulated by {\sc geant3}.  The TGC values are
chosen to be in the parameter space near the existing limits. For each
sample, the product of acceptance and efficiency is extracted from the
simulation as a ratio of the reconstructed and generated yields.
These ratios are averaged together as a function of $Z$ $p_T$ using
the maximum variation as an estimate of the uncertainty due to
assuming the efficiency and acceptance are not dependent on the TGC
values.

A likelihood for each of the couplings, $\mathcal{L}(\lambda_Z)$,
$\mathcal{L}(\Delta g_1^Z)$, and $\mathcal{L}(\Delta \kappa_Z)$, is
computed as a product of the Poisson probability of each of the bins
of the $Z$ $p_T$ distribution for the assumed anomalous coupling. Then 95\% confidence
levels are set where $(-2\text{ln}\mathcal{L}) -
(-2\text{ln}\mathcal{L}_{\text{min}}) = (1.96)^2$. 
The systematic
uncertainties include everything considered for the $WZ$ cross section
and the additional $p_T$-dependent uncertainty on the efficiency
which ranges from 5\% to 20\%.  
Systematic uncertainties are implemented in a way that most reduces the
TGC limit sensitivity when fluctuating the signal and background
by one standard deviation, thereby taking a conservative approach in
assigning systematic uncertainty.
The observed 95\% confidence level limits are consistent
with expectations as shown in Table \ref{tab:TGC}.

\begin{table}
\caption{Expected and observed limits on anomalous TGCs.  For each
  coupling limit set, the two other couplings are constrained at their SM
  values.  Values of the couplings outside the given observed range
  are excluded at the 95\% confidence level.}
\label{tab:TGC}
\begin{tabular}{llcccc}
\hline \hline
&&$ \Lambda$ (TeV)  &   $\lambda_Z$ &  $\Delta g_1^Z$  &   $\Delta \kappa_Z$ \\
\hline
Exp. & ~~ & 1.5     & ($-0.11,\ 0.12$) & ($-0.12,\ 0.23$) & ($-0.58,\ 0.94$) \\
Obs. & ~~ & 1.5     & ($-0.09,\ 0.11$) & ($-0.09,\ 0.22$) & ($-0.42,\ 0.99$) \\ 
\hline 	 
Exp. & ~~ & 2.0     & ($-0.10,\ 0.10$) & ($-0.11,\ 0.20$) & ($-0.53,\ 0.86$) \\
Obs. & ~~ & 2.0     & ($-0.08,\ 0.10$) & ($-0.08,\ 0.20$) & ($-0.39,\ 0.90$) \\
\hline \hline
\end{tabular}
\end{table}

To summarize, the $WZ$ production cross section has been measured in $p
\bar p$ collisions at $\sqrt{s} = 1.96$ TeV from reconstructed events
in the trilepton plus $\Met$ final state using a likelihood ratio
formed from a NeuroBayes neural network distribution that
discriminates signal from background.  This result, $\sigma (p \bar p \to WZ) =
3.93_{-0.53}^{+0.60}(\text{stat})_{-0.46}^{+0.59}(\text{syst}) $ pb, is the most precise
measurement at this energy with an overall uncertainty of less than
20\% and in agreement with SM predictions. 
The same event sample is also used to perform the most sensitive
probe to date at this energy of anomalous $WWZ$ couplings.  The $Z$
$p_T$ distribution of the sample is found to be in agreement with the
SM expectation and is used to place limits on anomalous triple gauge
couplings.

We thank the Fermilab staff and the technical staffs of the participating 
institutions for their vital contributions. This work was supported by the
U.S. Department of Energy and National Science Foundation; the Italian 
Istituto Nazionale di Fisica Nucleare; the Ministry of Education, Culture, 
Sports, Science and Technology of Japan; the Natural Sciences and Engineering 
Research Council of Canada; the National Science Council of the Republic of China; 
the Swiss National Science Foundation; the A.P. Sloan Foundation; 
the Bundesministerium f\"ur Bildung und Forschung, Germany; 
the Korean World Class University Program, the National Research Foundation of Korea; 
the Science and Technology Facilities Council and the Royal Society, UK; 
the Institut National de Physique Nucleaire et Physique des Particules/CNRS; 
the Russian Foundation for Basic Research; the Ministerio de Ciencia e Innovaci\'{o}n, 
and Programa Consolider-Ingenio 2010, Spain; the Slovak R\&D Agency; the Academy of Finland; 
and the Australian Research Council (ARC).


\end{document}